\documentclass[twocolumn,letterpaper]{IEEEtran}

\usepackage[
    top=0.7in,
    bottom=1.05in,
    left=0.59in,
    right=0.59in
]{geometry}

\usepackage{lipsum}
\usepackage{amsmath}
\usepackage{amsfonts}
\usepackage{amssymb}
\usepackage[ruled,vlined]{algorithm2e}
\usepackage{caption}
\usepackage{mathtools}  
\usepackage{amsthm}
\usepackage{booktabs}
\usepackage{subfigure}
\usepackage{subcaption}
\usepackage{algpseudocode}
\usepackage{tabulary}
\usepackage{footnote}
\usepackage[export]{adjustbox}
\usepackage{booktabs}
\usepackage[justification=centering]{caption}
\usepackage{comment}
\newtheorem{remark}{Remark}
\usepackage{cases}
\usepackage{graphicx}
\usepackage{cite}
\usepackage{multicol}
\usepackage{xcolor}
\usepackage{graphicx}
\graphicspath{{./}{figures/}}
\newcommand\numeq[1]%
  {\stackrel{\scriptscriptstyle(\mkern-1.5mu#1\mkern-1.5mu)}{=}}
\usepackage{stfloats}
\usepackage{cuted}
\usepackage{lipsum}  

\usepackage{subfigure}

\title{Optimized Non-Primary Channel Access Design in IEEE 802.11bn}
\author{\text{Dongyu Wei},  \text{Liu Cao}, \text{Lyutianyang Zhang}, \text{Xiangyu Gao}, \text{Hao Yin}
\thanks{Dongyu Wei, Liu Cao, Lyutianyang Zhang, Xiangyu Gao and Hao Yin are with the Department of Electrical \& Computer Engineering, University of Washington, Seattle, WA, USA (e-mail:\{wdy1223, liucao, haoyin, lyutiz, xygao\}@uw.edu).}}%

\begin{document}

\pagestyle{empty}

\maketitle
\thispagestyle{empty}

\begin{abstract}
The IEEE 802.11 standards, culminating in IEEE 802.11be (Wi-Fi 7), have significantly expanded bandwidth capacities from 20 MHz to 320 MHz, marking a crucial evolution in wireless access technology. Despite these advancements, the full potential of these capacities remains largely untapped due to inefficiencies in channel management, in particular, the underutilization of secondary (non-primary) channels when the primary channel is occupied. This paper delves into the Non-Primary Channel Access (NPCA) protocol, initially proposed by the IEEE 802.11 Ultra-High Reliability (UHR) group, aimed at addressing these inefficiencies. Our research not only proposes an analytical model to assess the throughput of NPCA in terms of average throughput but also crucially identifies that the overhead associated with the NPCA protocol is significant and cannot be ignored. This overhead often undermines the effectiveness of the NPCA, challenging the assumption that it is invariably superior to traditional models. Based on these findings, we have developed and simulated a new hybrid model that dynamically integrates the strengths of both legacy and NPCA models. This model overall outperforms the existing models under all channel occupancy conditions, offering a robust solution to enhance throughput efficiency.
\end{abstract}

\section{Introduction} 
\label{introduction}
The IEEE 802.11 standards, developed by the Institute of Electrical and Electronics Engineers (IEEE), form the bedrock of modern wireless network technologies, finding diverse applications across numerous fields. Historically, these standards have seen considerable evolution in terms of bandwidth capacity, escalating from an initial 20 MHz in IEEE 802.11a to a robust 160 MHz in the later IEEE 802.11ac and IEEE 802.11ax iterations  \cite{masiukiewicz2019throughput}. The advent of IEEE 802.11be, also known as the Extremely High Throughput (EHT) standard, marks a monumental stride in this progression, introducing support for up to 320 MHz bandwidth \cite{lopez2019ieee, hoefel2020ieee, 9152055}. This significant expansion in bandwidth capability necessitates a reevaluation of channel allocation methodologies. These approaches supports the use of secondary channels for stations (STAs) when interfaced with an access point (AP) capable of harnessing the wide bandwidth. This strategy effectively reduces the overload on primary channels by stations operating at lower bandwidths, thereby enhancing the overall efficiency and distribution of bandwidth across both primary and secondary channels.

While the IEEE 802.11 standards have made considerable strides in enhancing bandwidth capacities, each iteration has consistently incorporated a primary 20 MHz channel to ensure backward compatibility with older versions. This requirement for a consistent primary channel, though crucial for interoperability, occasionally hampers overall system performance. This is because the compulsory use of the primary channel for data transmission can restrict the operational flexibility of the secondary channels, which may remain idle during periods of primary channel occupied activity. Despite the forward-thinking designs of the IEEE 802.11be or EHT standard, which promotes secondary channel communications \cite{8413100, 10500831}, there are still significant challenges that need to be addressed, particularly regarding how access points (APs) manage and initiate data transmission on secondary channels during times when the primary channel is saturated. In response to these challenges, the IEEE 802.11 Ultra-High Reliability (UHR) group has introduced a proposal for Non-Primary Channel Access (NPCA) \cite{IEEE80211-23/0797r0, IEEE80211-24/0486r0}. This innovative approach aims to enhance the utilization efficiency of channels by enabling more dynamic employment of non-primary channels under congested conditions.

Extensive research has been conducted on the throughput performance determined by the IEEE 802.11 standard channel access \cite{lopez2019ieee, bianchi2000performance, parker2015increasing, masiukiewicz2019throughput, yang2020survey, khorov2020current, galati2023will, 10488297, zhang2023ieee, lopez2022multi, 9903472}. Among these studies, the Distributed Coordination Function (DCF) network model proposed by Bianchi in \cite{bianchi2000performance} was extreme influential. This model utilizes a two-dimensional Markov chain to represent a saturated network environment, providing a robust framework for analyzing throughput in the Base Station Subsystem (BSS). The simplicity and effectiveness of Bianchi's model have spurred various extensions and improvements \cite{parker2015increasing, masiukiewicz2019throughput}, focusing primarily on enhancing throughput analysis. Further advancements have been addressed under the next generation of IEEE standards, including IEEE 802.11be (Wi-Fi 7) \cite{yang2020survey, khorov2020current} and the upcoming IEEE 802.11bn known as Wi-Fi 8 or Ultra High Rate (UHR) \cite{galati2023will, 10488297}. These standards introduce capabilities such as increased bandwidth up to 320 MHz, Multi-Link Operation (MLO), and multi-band/multi-channel aggregation, which promises further throughput optimizations, leveraging advanced technologies to accommodate the growing demand for faster and more reliable wireless communication systems \cite{zhang2023ieee, lopez2022multi, 9903472}.

However, the historical work on the NPCA protocol, as proposed by the IEEE 802.11 UHR group, still requires further consideration and analysis in throughput optimization. While shifting to non-primary channels promises enhanced performance, we find that the transition overhead and associated delays cannot be overlooked. Our studies indicate that the overhead from frequent channel switching may reduce NPCA system efficiency, potentially making it less effective than traditional networks due to the significant time costs involved. This paper not only makes comparison and discussion over the NPCA and legacy models, but also proposes a new model that blends the strengths of both approaches, aiming to improve overall throughput across all the channel conditions. Key contributions of this paper include:
\begin{itemize}
    \item The development of a new analytical model for a two-channel system, designed to evaluate the throughput of NPCA networks.
    \item Incorporation of overhead considerations into the model to compare NPCA with legacy networks across various channel occupancy scenarios.
    \item Performance simulations of both NPCA and legacy networks, leading to the creation of a dynamic model that surpasses existing models in throughput efficiency.
\end{itemize}

The structure of this paper is as follows: Section II discusses traditional Bianchi's model and the NPCA mechanism. Section III details the analytical model and throughput analysis within NPCA networks. Section IV presents validation simulations and introduces our newly proposed model. The paper concludes in Section V.

\vspace{-3mm}
\section{NPCA Network Operations}
\label{sec:sys_arc}
\begin{figure}[tp]
    \centering
\includegraphics[width=.4\textwidth]{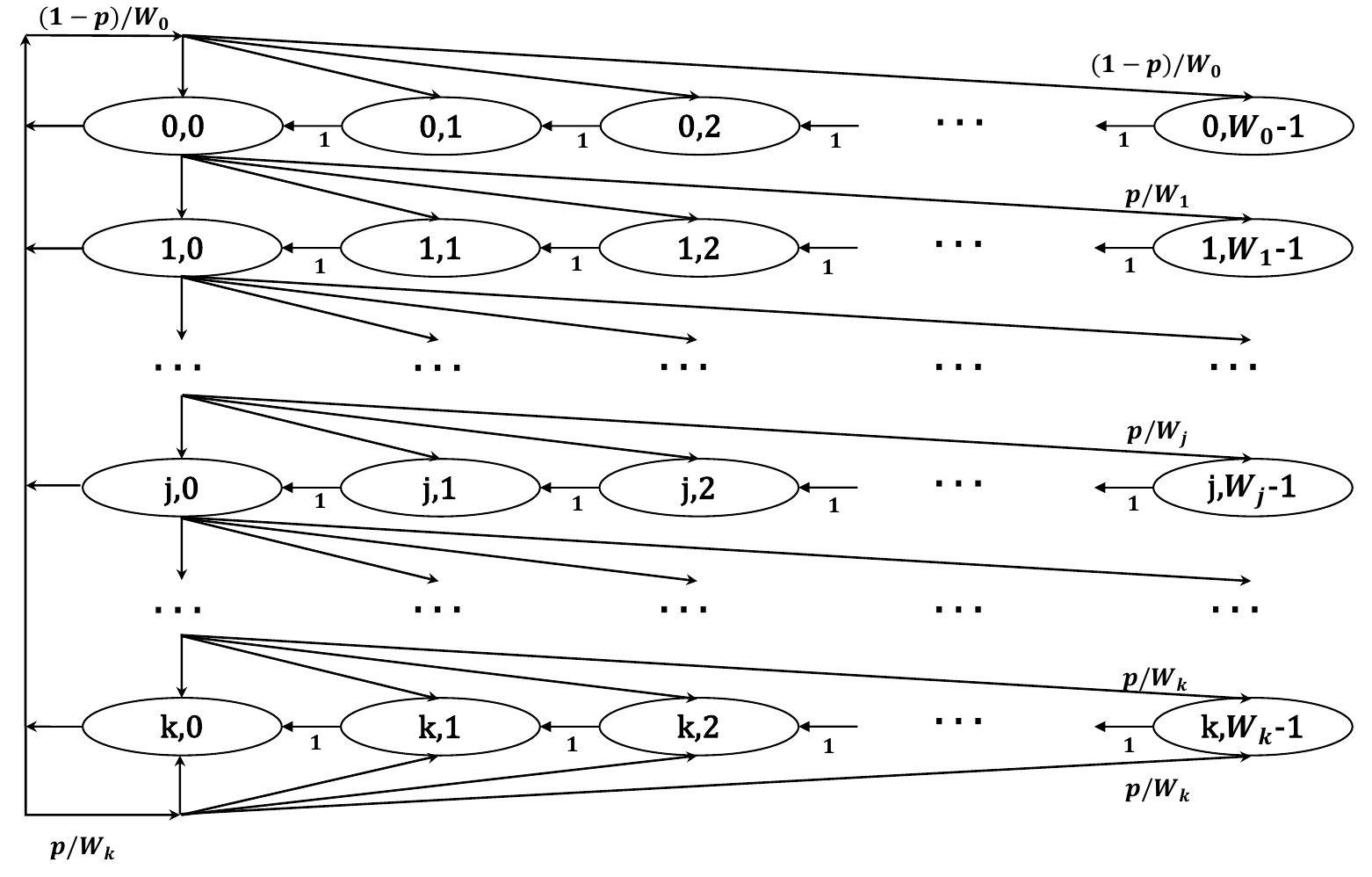}
    \caption{Markov Chain Model of Contention Window Backoff in CSMA/CA.}
    \label{fig:markov}
    \vspace{-0.2cm}
\end{figure}

In the IEEE 802.11 standards, wireless devices utilize the Carrier Sense Multiple Access with Collision Avoidance (CSMA/CA) protocol. A wireless device attempting to transmit on its primary channel first senses and waits until the channel is idle for a Distributed Interframe Space (DIFS). As shown in Fig. \ref{fig:markov}, during the backoff process, the device randomly selects an integer value from $[0, W-1]$ as the backoff counter, where $W$ is the contention window size. The counter decrements by one at the end of each idle slot. If the channel becomes busy, the decrement is paused until the channel is idle again for a DIFS period. Upon the backoff counter reaching zero, the device initiates transmission. If the transmission fails, the contention window size $W$ doubles, and the process repeats up to $k$ maximum retransmission attempts.

\begin{figure}[ht]
\centering
 \subfigure[The Backoff Process in Legacy Network.]
{\includegraphics[width=0.4\textwidth]{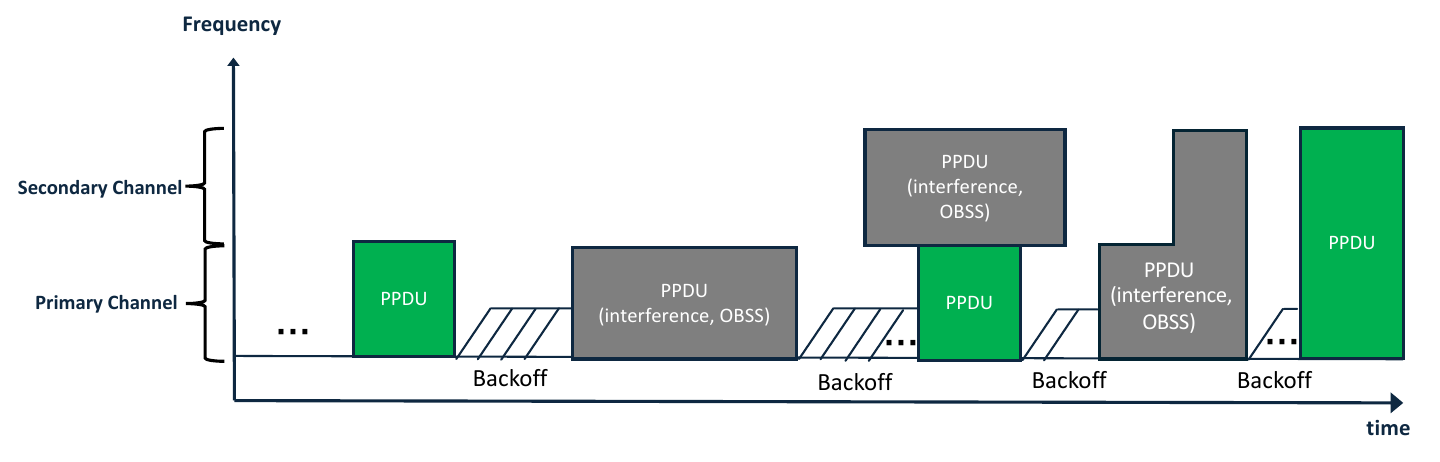}}
\label{fig:2a}
\centering \\
\subfigure[The Backoff Process in NPCA Network.]{
\includegraphics[width=0.4\textwidth]{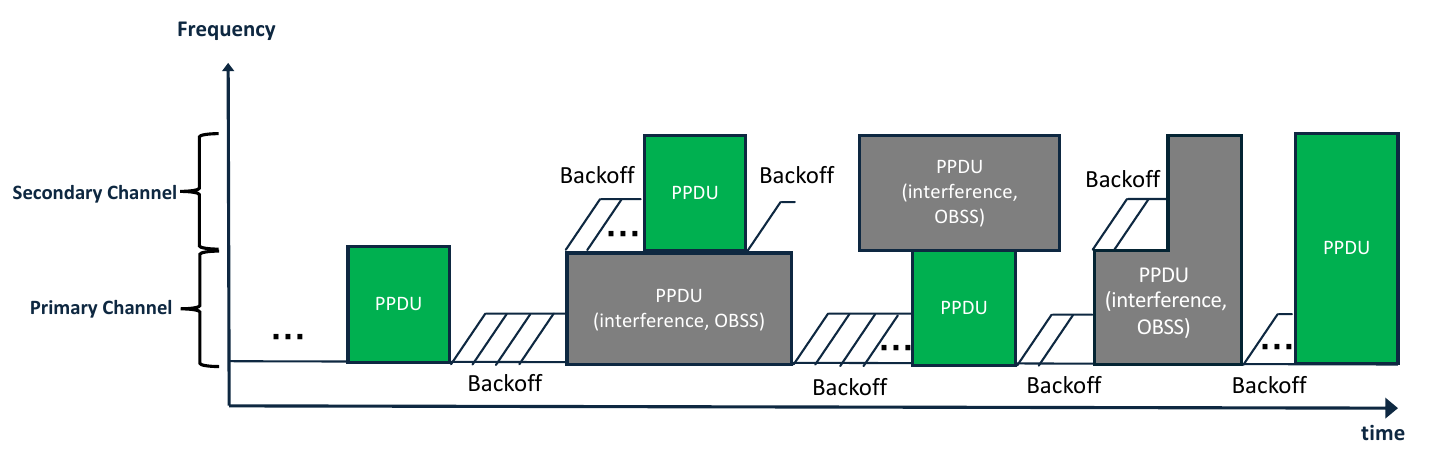}}
\label{fig:2b}
\caption{The Transmission Process Comparison between Legacy and NPCA Network.}
\label{fig:LegacyvsNPCA}
\vspace{-0.2cm}
\end{figure}

The legacy devices, as depicted in Fig. \ref{fig:LegacyvsNPCA}(a), can transmit on the primary channel and also utilize an idle secondary channel to achieve higher throughput, but only if the primary channel is idle. This dependency on the primary channel's status underutilizes the entire channel system. In contrast, the NPCA model, as illustrated in Fig. \ref{fig:LegacyvsNPCA}(b), enhances performance by switching to an idle non-primary channel when the primary one is busy. In this example, after the first transmission, when the NPCA network senses Physical Protocol Data Units (PPDUs) from Overlapping BSS (OBSS) on the primary channel, it will switch to the secondary channel if it is sensed idle. The backoff counter is carried over to the secondary channel during channel switch so that the network can transmit more PPDUs. When the primary channel becomes idle during the backoff process, the network switches back. This performance enhances the channel efficiency of the network by using more available channels during transmission and increasing throughput.

\begin{figure}[htp]
    \centering
\includegraphics[width=.4\textwidth]{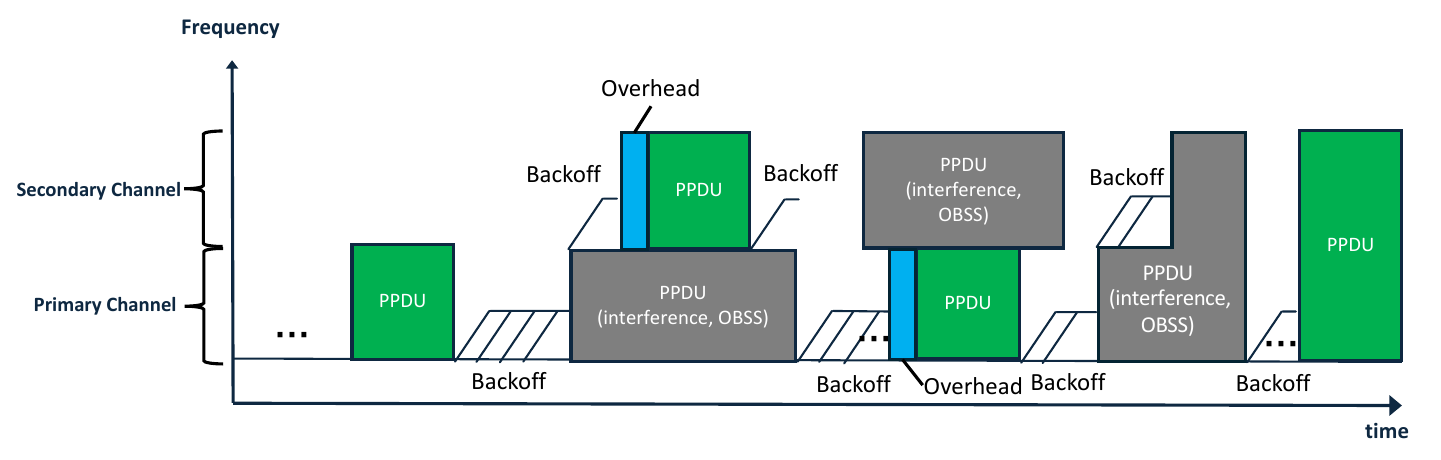}
    \caption{The Two-Channel NPCA Model with Overhead.}
    \label{fig:overhead}
\end{figure}

However, the classic NPCA model does not account for the overhead effects, which significantly impact its efficiency. As shown in Fig. \ref{fig:overhead}, switching channels incurs a delay, as the device may need to synchronize by sending control frame when changing the transmission channel. This overhead, though has not been finalized yet in 802.11 UHR protocol, can potentially go up to several milliseconds that degrades the NPCA model's throughput \cite{6898027}, especially with frequent channel switch. This issue necessitates a trade-off consideration between overhead delays and increased transmission opportunities on multiple channels. 

\vspace{-3mm}
\section{Analytical model}
\label{sec:sys_model}
To delve deeper into the model analysis, we first consider a scenario within a BSS where $n$ nodes perform transmissions over a single-channel network. Drawing from Bianchi's model \cite{bianchi2000performance}, we denote $S$ as the throughput of the single channel, determined by $P_{tr}$, the probability that one or more nodes transmit in a given slot time. The probability $\tau$ of a station ready to transmit in a random slot is defined as:
\begin{equation}
    P_{tr} = 1 - (1 - \tau)^{n}. \label{eq:ptr}
\end{equation}
The probability $P_s$ of a successful transmission by exactly one station on the channel is expressed as:
\begin{equation}
  P_{s} = \frac{n\tau(1-\tau)^{n-1}}{P_{tr}} = \frac{n\tau(1-\tau)^{n-1}}{1 - (1 - \tau)^{n}}. \label{eq:ps}
\end{equation}
The average throughput $S$ can then be expressed as the ratio:
\begin{equation}
    S = \frac{E[\text{payload information transmitted in a slot time}]}{E[\text{length of a slot time}]}, \label{eq:s1}
\end{equation}
with $\sigma$ representing the duration of an empty slot, and $E[P]$ the average packet payload size. It is further refined by:
\begin{equation}
  S = \frac{P_s P_{tr} E[P]}{(1 - P_{tr})\sigma + P_{tr} P_s T_s + P_{tr}(1-P_s)T_c}, \label{eq:s}
\end{equation}
where $T_s$ and $T_c$ are the average times the channel is busy for a successful transmission and a collision, respectively. These times include transmission delays such as headers, acknowledgments, and interframe spaces:
\begin{flalign}
  T_{s} & = H + E[Pkt] + \text{SIFS} + \delta + \text{ACK} + \text{DIFS} + \delta, \label{eq:ts}&&\\
  T_{c} & = H + E[Pkt] + \delta + \text{EIFS}, \label{eq:tstc}&&
\end{flalign}
where $H$ is the packet header ($H = PHY_{hdr} + MAC_{hdr}$), and $\text{SIFS}$, $\text{DIFS}$, and $\text{EIFS} = \text{SIFS} + \text{NACK} + \text{DIFS}$ are time durations critical for processing frames and reducing further collisions.

Next, we extend the analytical model from a single channel to a two-channel system, where we analyze the throughput of both legacy and NPCA networks. As shown in Fig. \ref{fig:LegacyvsNPCA}, we denote the primary channel as channel 1 and the non-primary as channel 2, with the rate occupied by OBSS $p_1$ and $p_2$. In the legacy network, as shown in Fig. \ref{fig:LegacyvsNPCA}(a), the throughput on channel 1, $Th_1$, is $S(p_1)$, dependent solely on this channel being idle. As revealed in Fig. \ref{fig:LegacyvsNPCA}(b), if channel 1 is idle, the throughput on channel 2, $Th_2$, is given by:
\begin{equation}
  Th_2 = Th_1(1 - p_2) = S(p_1)(1 - p_2), \label{eq:th2_leg}
\end{equation}
resulting in a total throughput for the two-channel legacy network:
\begin{equation}
  S_{leg} = Th_1 + Th_2 = S(p_1)(2 - p_2). \label{eq:s_leg}
\end{equation}

For NPCA model, we first analyze the classic model from the IEEE 802.11 UHR group, which does not include overhead. Here, the whole throughput the model transmitting on primary channel (channel 1) is $W_1 = S_{leg} = S(p_1)(2 - p_2)$. If channel 1 is busy, the device may switch to channel 2 if it is idle, with the throughput given by:
\begin{equation}
  W_{2} = S(p_1)\left(\frac{p_1}{1 - p_1}\right)(1 - p_2), \label{eq:th1}
\end{equation}
combining to form the total throughput for the classic NPCA network:
\begin{equation}
  S_{npca}^* = W_1 + W_2 = S(p_1)[(2 - p_2) + \left(\frac{p_1}{1 - p_1}\right)(1 - p_2)]. \label{eq:s_npca}
\end{equation}
\begin{remark}
As per (\ref{eq:s_leg}) and (\ref{eq:s_npca}), it is evident that $S_{leg} = S(p_1)(2 - p_2)$, whereas $S_{npca}^* = S_{leg} + S(p_1)\left(\frac{p_1}{1 - p_1}\right)(1 - p_2)$, demonstrating that $S_{npca}^* > S_{leg}$, confirming the superior throughput of NPCA over the legacy network under ideal conditions without overhead.
\end{remark}

\begin{figure}[htp]
    \centering
    \includegraphics[width=.3\textwidth]{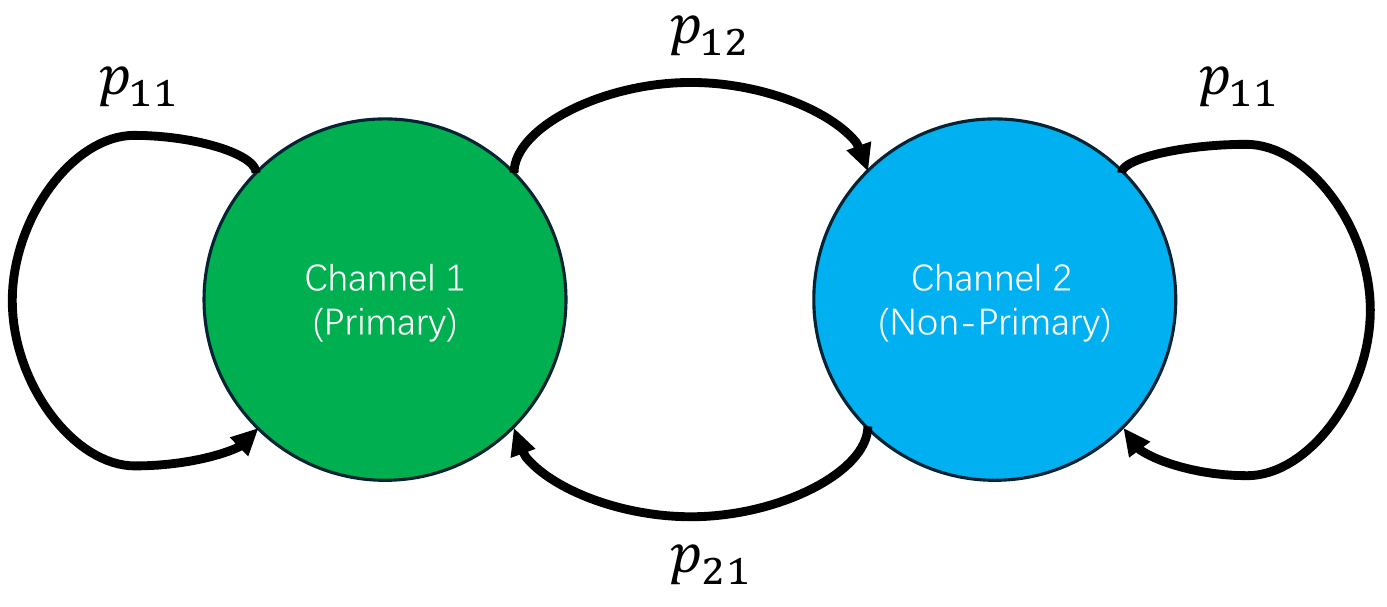}
    \caption{The State Machine Transition Probability Distribution.}
    \label{fig:machine}
\end{figure}

However, as depicted in Fig. \ref{fig:overhead}, incorporating the overhead into the model makes the output complicated but more precise and practical. As shown in Fig. \ref{fig:machine}, the model illustrates the current channel state on which the device is transmitting and the conditional transition probability for the next transmission. The probability $p_{ij}$ represents the transition probability of the device switching from channel $i$ to channel $j$ in the next transmission slot. From Table \ref{tab:set}, we derive that during each transmission process, the probability that the device transmits on channel 1 is given by:
\begin{equation}
  P_{tr}^1 = \frac{1 - p_1}{1 - p_1p_2}, \label{eq:p_tr_1}
\end{equation}
and the probability that the device transmits on channel 2 is expressed as:
\begin{equation}
  P_{tr}^2 = \frac{p_1 - p_1p_2}{1 - p_1p_2}. \label{eq:p_tr_2}
\end{equation}

\begin{table}[ht]
\centering 
\begin{tabular}{|c|c|c|c|c|} 
\hline 
Ch2 & Occupied & Occupied & Idle & Idle \\ \hline 
Ch1 & Occupied & Idle & Occupied & Idle \\ \hline 
Prob & $p_1p_2$ & $(1-p_1)p_2$ & $p_1(1-p_2)$ & $(1-p_1)(1-p_2)$ \\ \hline 
\end{tabular}
\caption{Event Set of Two-Channel Model} 
\label{tab:set} 
\end{table}

Given that channel states in any time slot are independent events, the transition matrix $T$ shown in Fig. \ref{fig:machine} can be constructed as follows:
\begin{equation}
T = \begin{pmatrix}
    P_{tr}^1 & P_{tr}^1 \\
    P_{tr}^2 & P_{tr}^2
\end{pmatrix}
= \begin{pmatrix}
    \frac{1 - p_1}{1 - p_1p_2} & \frac{1 - p_1}{1 - p_1p_2} \\
    \frac{p_1 - p_1p_2}{1 - p_1p_2} & \frac{p_1 - p_1p_2}{1 - p_1p_2}
\end{pmatrix}.\label{eq:matrix}
\end{equation}
We assign the probability of the device transmitting on each channel as $Pb_1$ and $Pb_2$. During any time slot $t$, the probability distribution matrix of the device transmitting on different channels can be given by:
\begin{equation}
P(t) = \begin{pmatrix}
    Pb_1  \\
    Pb_2
\end{pmatrix}
\label{eq:matrix_distribution}.
\end{equation}
The initial state $P(0)$ is 
$\begin{pmatrix}
    1  \\
    0
\end{pmatrix}$,
indicating that channel 1 is the primary channel. When the system reaches a steady state, it should satisfy:
\begin{equation}
T \cdot P(t) = P(t+1), \quad Pb_1 + Pb_2 = 1
\label{eq:steady_state}.
\end{equation}
The steady-state distribution of the device transmitting on channels 1 and 2 can be expressed as:
\begin{gather}
  Pb_1 = \frac{1 - p_1}{1 - p_1p_2}, \label{eq:pb1}\\
  Pb_2 = \frac{p_1 - p_1p_2}{1 - p_1p_2}. \label{eq:pb2}
\end{gather}
In any time slot, as shown in Table \ref{tab:ov}, if the channel the device is currently going to transmit on differs from that of the last transmission, it will incur overhead, which is the delay for the device to switch channels and broadcast synchronization signals. The probability of incurring overhead for devices transmitting on channel 1 is:
\begin{equation}
Po_1 = \frac{Pb_2}{Pb_1 + Pb_2}
\label{eq:po1},
\end{equation}
and similarly, the probability for channel 2 is:
\begin{equation}
Po_2 = \frac{Pb_1}{Pb_1 + Pb_2}
\label{eq:po2}.
\end{equation}
From Eq. \ref{eq:s_npca}, we obtain $W_1$ and $W_2$, and each must be adjusted by a coefficient considering the overhead. These coefficients for channels 1 and 2 can be calculated as:
\begin{gather}
  c_1 = \frac{Po_1 + Po_2}{l \cdot Po_2 + Po_1}, \label{eq:c1}\\
  c_2 = \frac{Po_1 + Po_2}{l \cdot Po_1 + Po_2}, \label{eq:c2}
\end{gather}
where $l = \frac{\text{Length of PPDU + Length of Overhead}}{\text{Length of PPDU}}$. Thus, the throughput of the NPCA considering overhead can be derived as:
\begin{equation}
\begin{aligned}
S_{npca} &= c_1 \cdot W_1 + c_2 \cdot W_2 \\
         &= S(p_1)\{\frac{(1-p_1p_2)(2-p_2)}{lp_1(1-p_2)+1-p_1} \\
         &+ \frac{(1-p_1p_2)p_1(1-p_2)}{[p_1(1-p_2)+l(1-p_1)](1-p_1)}\}
\end{aligned}
\label{eq:real_npca}
\end{equation}

\begin{table}[ht]
\centering 
\begin{tabular}{|c|c|c|c|} 
\hline 
Last Transmission & Current Transmission & Probability & Overhead \\ \hline 
Ch1 & Ch1 & $Pb_1 \cdot Pb_1$ & No \\ \hline 
Ch1 & Ch2 & $Pb_1 \cdot Pb_2$ & Yes \\ \hline 
Ch2 & Ch1 & $Pb_2 \cdot Pb_1$ & Yes \\ \hline
Ch2 & Ch2 & $Pb_2 \cdot Pb_2$ & No \\ \hline
\end{tabular}
\caption{Event Set of Two-Channel Model} 
\label{tab:ov} 
\end{table}

\vspace{-5mm}
\section{Simulation}
This section evaluates the throughput performance of NPCA networks compared to legacy networks. Using Eqs. \ref{eq:real_npca} and \ref{eq:s_leg}, we express the throughput ratio of NPCA to legacy networks as:
\begin{equation}
\begin{aligned}
  \frac{S_{npca}}{S_{leg}} &= \frac{1-p_1p_2}{lp_1(1-p_2)+1-p_1}\\ 
  &+ \frac{1-p_1p_2}{p_1(1-p_2)+l(1-p_1)} \cdot \frac{p_1}{1-p_1} \cdot \frac{1-p_2}{2-p_2}. \label{eq:ratio}
\end{aligned}
\end{equation}
Here, when $\frac{S_{npca}}{S_{leg}} > 1$, NPCA demonstrates superior throughput; otherwise, the legacy network outperforms NPCA in throughput efficiency.

We simulate a single-BSS scenario with ten nodes that sense and contend for transmission using Bianchi's rule. The legacy network is restricted to the primary channel, while NPCA can switch to a non-primary channel when the current channel is busy. However, channel switching introduces overhead, controlled by the parameter \( l \) in Eq. \ref{eq:real_npca}, reflecting hardware variability in overhead magnitude. We establish three distinct simulation scenarios to systematically compare NPCA and legacy network throughputs under varying channel and network conditions. Due to significant overhead we discussed, it is crucial to maintain $l$ within a reasonable range. In our simulation, we utilize the Aggregate MAC Protocol Data Unit (AMPDU) for data transmission. AMPDU allows the bundling of multiple MAC Protocol Data Units (MPDUs) into a single large packet, effectively reducing the overhead \cite{ong2011ieee}. This aggregation mechanism is anticipated to be increasingly prevalent in the context of 802.11be and Ultra-High Rate (UHR) scenarios to maximize overall throughput and improve channel efficiency \cite{mao2023network}. The simulation parameters are summarized in Table \ref{tb:simu_para}.

\begin{table}[htp]
\caption{Simulation Parameters}
\label{tb:simu_para}
\centering
\fontsize{8}{8}\selectfont{
\begin{tabular}{|c|c|}
\hline
\textbf{Parameters} & \textbf{Value} \\ \hline
Simulation Time (s) & 30 \\ \hline
Number of Channels & 2 \\ \hline
Number of BSS & 1 \\ \hline
Number of Stations per BSS & 10 \\ \hline
Primary Channel & Channel 1 \\ \hline
Channel Bandwidth (MHz) & 20 \\ \hline
Channel Utilization & 1 \\ \hline
Packet Size (Bytes) & 1500 \\ \hline
AMPDU Size (Bytes) & 18000 \\ \hline
$l$ & 1.8, 2.0, 2.2 \\ \hline
MCS & 3 \\ \hline
CW Min, $CW_{min}$ & 16 \\ \hline
CW Max, $CW_{max}$ & 1024 \\ \hline
Slot (us) & 9 \\ \hline
SIFS (us) & 16 \\ \hline
DIFS (us) & SIFS + 2 $\cdot$ Slot \\ \hline
\end{tabular}
}
\end{table}

Each scenario aims to test the average throughput of the NPCA and legacy models to analyze and provide a comprehensive study of their performance across various operational conditions.

\begin{figure*}[t]
\begin{minipage}[t]{0.32\linewidth}
\centering
 \subfigure[$p_1$ is high($p_1 = 0.8$) while $p_2$ is low.]
{\includegraphics[width=2in]{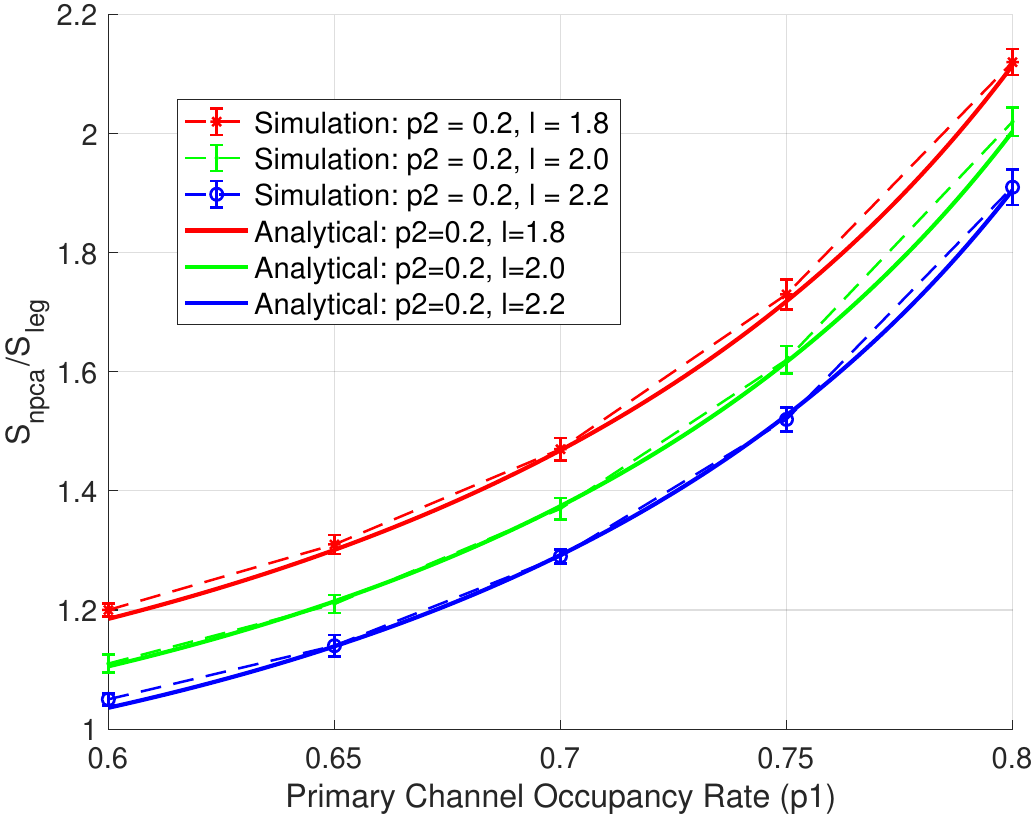}}
\label{fig:p1_h}
\end{minipage}
\begin{minipage}[t]{0.32\linewidth}
\centering
\subfigure[$p_1$ is low while $p_2$ is high($p_2 = 0.8$).]{
\includegraphics[width=2in]{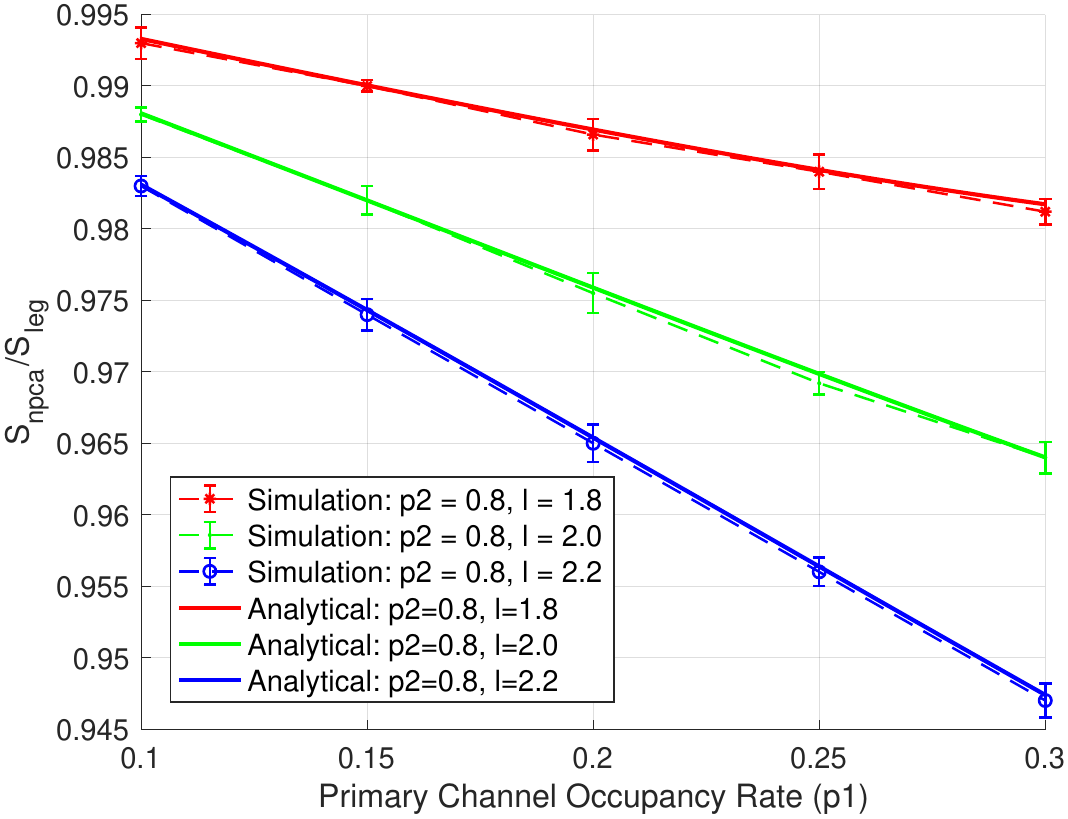}}
\label{fig:p2_h}
\end{minipage}%
\begin{minipage}[t]{0.32\linewidth}
\centering
\subfigure[$p_1$ is equal to $p_2$.]{
\includegraphics[width=2in]{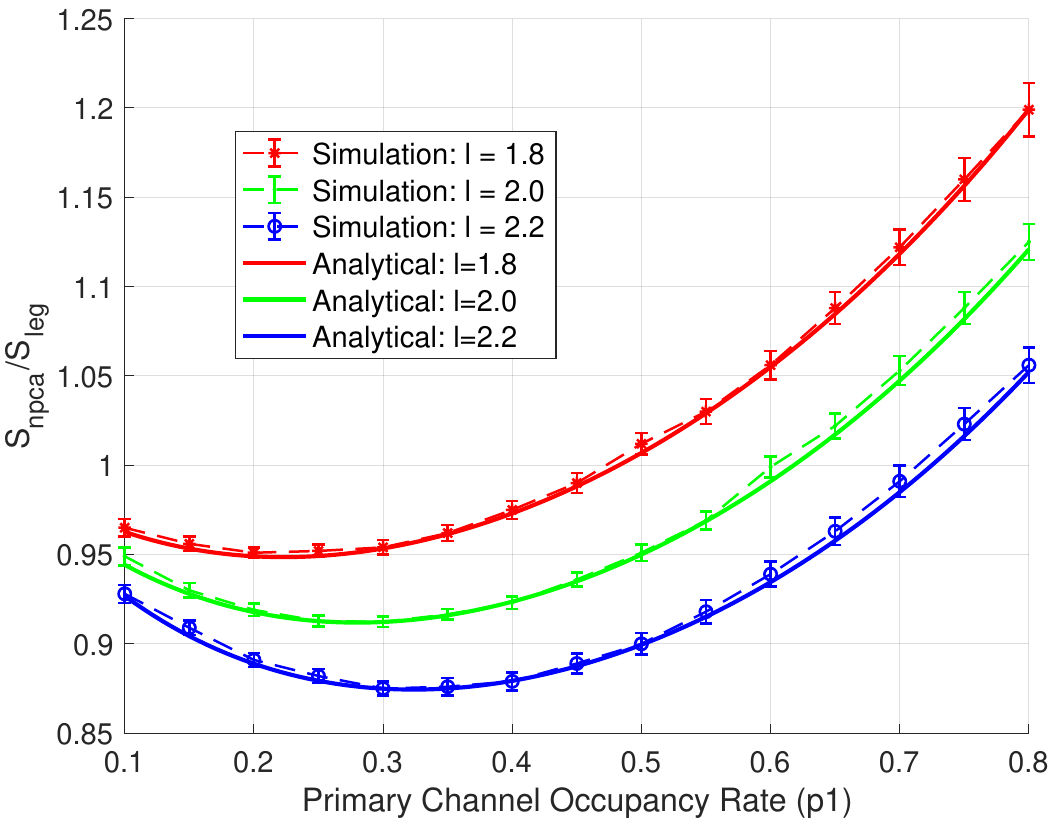}}
\label{fig:p1_p2}
\end{minipage}
\caption{The ratio of throughput between the legacy network and the NPCA network.}
\label{fig:AoI_sim}
\end{figure*}
\vspace{-3mm}
\subsection{\(p_1\) is high while \(p_2\) is low}
When the primary channel is highly occupied (\(p_1 > 0.6\)) and the non-primary channel is relatively largely idle (\(p_2 < 0.3\)), from Eq. \ref{eq:ratio} we predict that \(\frac{S_{npca}}{S_{leg}} \approx \frac{1}{l}+\frac{p_1}{1-p_1} > 1\). In this scenario, the legacy network is likely to encounter collisions due to its reliance on the idle primary channel for transmissions. Conversely, the NPCA network, despite delays from frequent channel switching, can exploit the idle non-primary channel, potentially achieving higher throughput. As depicted in Fig. \ref{fig:AoI_sim}(a), with \(p_2 = 0.2\) and \(p_1\) varying from 0.6 to 0.8, the busier the primary channel, the more advantageous NPCA becomes. Different values of \(l\) demonstrate how reducing overhead can further enhance NPCA's performance.
\vspace{-3mm}
\subsection{\(p_1\) is low while \(p_2\) is high}
In scenarios where the primary channel is idle (\(p_1 < 0.3\)) and the non-primary channel is busy (\(p_2 > 0.6\)), as illustrated in Fig. \ref{fig:AoI_sim}(b), with \(p_2 = 0.8\) and \(p_1\) between 0.1 and 0.3, the ratio \(\frac{S_{npca}}{S_{leg}} \approx 1-p_1p_2 < 1\). This finding aligns with our simulation results. Here, NPCA's channel switching incurs significant overhead, and transmissions on the busy non-primary channel are prone to collisions or frequent occupies by OBSS, diminishing the benefits of NPCA compared to the legacy network.
\vspace{-5mm}
\subsection{\(p_1 \approx p_2\)}
For similar occupancy rates in both channels, we denote $p_1 = p_2 = p$ to simplify the model, and the throughput ratio is described by:
\begin{equation}
  \frac{S_{npca}}{S_{leg}} = \frac{p+1}{lp+1} + p \cdot \frac{p+1}{(l+p)(2-p)}. \label{eq:balanced_ratio}
\end{equation}
As shown in Fig. \ref{fig:AoI_sim}(c), low occupancy rates allow both networks ample transmission opportunities, making NPCA's channel switching less beneficial or even worse than legacy network due to the cost on switching channels. Conversely, high occupancy rates mirror the conditions where NPCA's ability to switch channels gets more chances to transmit and improves the throughput. There exists a threshold where the ratio equals 1, delineating the conditions under which each model performs better.
\vspace{-3mm}
\subsection{Discussion and Advanced Model}
The simulation results indicate that, contrary to the  initial findings from IEEE 802.11 UHR group, NPCA does not always outperform the legacy network, particularly considering overhead and channel conditions. The decision between using NPCA or legacy networks heavily depends on the occupancy rate of the primary channel (\(p_1\)). 

In scenarios where the primary channel (\(p_1\)) is lightly loaded, the legacy network tends to perform better. This is because the lower occupancy rate of the primary channel means that most packets can be transmitted with minimal delay, reducing the necessity for switching to the non-primary channel, which incurs additional overhead. Conversely, when \(p_1\) is high, indicating a busy primary channel, the NPCA network is more advantageous as it allows packets to be diverted to a less congested non-primary channel, thereby increasing the likelihood of successful transmissions without facing delays due to heavy channel contention.

To address the variability in performance based on channel conditions, we developed an advanced dynamic switching model that intelligently chooses between the NPCA and legacy models based on real-time channel occupancy assessments. This model is detailed in Algorithm~\ref{alg:alg1} and involves setting a threshold (\(thre_1\)), which determines the occupancy rate at which the switch to the legacy or to the NPCA model is triggered. The threshold \(thre_1\) and the look-back period \(k_1\) for assessing channel occupancy are configurable parameters that can be optimized based on the length of overhead the specific network costs and other transmission conditions or requirements.

This dynamic model operates by continuously monitoring the occupancy rate of the primary channel within the last \(k_1\) time slots. Based on this monitored data, the model dynamically decides whether to use the NPCA method or revert to the legacy method for each new time slot. The decision is based on whether the measured occupancy rate exceeds the set threshold \(thre_1\). By adaptively switching between the two models, this advanced system aims to optimize throughput by leveraging the strengths of both NPCA and legacy channel access methods under varying network load conditions.

The performance of this advanced model was simulated and compared against the standalone NPCA and legacy models under a range of channel occupancy scenarios. We designed a experiment, where we set the overhead relatively large at $l = 2.2$. We created the dynamical two-channel scenario, in which every period($1 \text{second}$), the two channels randomly select their occupancy rate $p$ to be idle ($p \in [0.1, 0.35)$), medium ($p \in [0.35, 0.6)$), or busy ($p \in [0.6, 0.85)$). We simulated 1000 periods and calculated the average throughput for each model during the whole process. As shown in Table \ref{tb:simu_out}, in this scenario where the occupancy rates of the two channels are randomly and uniformly distributed, the throughput of NPCA does not improve a lot, but the throughput of our model outperforms any other model for at least 10\%. The results indicate that the dynamic model overall is consistently greater than any standard model, proving its effectiveness in utilizing the most advantageous channel access method based on real-time traffic conditions.

\begin{algorithm}[tp]
\caption{Dynamic Channel Sensing and Model Utilization Algorithm.}
\label{alg:alg1}

Initialize: Ch1(Primary channel), Ch2(Non-primary channel)

Initialize parameters $p_1, thre_1, k_1$ for network model

\While{True} {
    \For{each time slot} {
        sense if Ch1 is busy or idle;
        
        refresh $p_1$ of Ch1 over latest $\max(1, k_1)$ time slots;

        \If{$p_1 > thre_1$} {
            Use NPCA method;
        }\Else {
            \If{Node is at Ch2} {
                Switch to Ch1;
            }
            Use Legacy method;
        }
    }
}

\end{algorithm}
\label{simulation}

\begin{table}[tp]
\caption{Simulation Outputs}
\label{tb:simu_out}
\centering
\begin{tabular}{|c|c|}
\hline
\textbf{Model} & \textbf{Throughput(Mbps)} \\ \hline
Legacy & 9.2575 \\ \hline
NPCA  & 9.5250 \\ \hline
Our model & 10.7251 \\ \hline
\end{tabular}
\end{table}
\vspace{-3mm}
\section{Conclusion}
\label{sec.conclusion}
This paper provided a detailed analysis of the throughput capabilities of the Non-Primary Channel Access (NPCA) protocol as proposed by the IEEE 802.11 UHR task group. We developed and validated analytical models for both the NPCA and the legacy network, conducting simulations to test our theoretical predictions. Our findings indicate that while NPCA can enhance throughput under certain conditions, its performance superiority is not consistent across all scenarios due to the overhead associated with channel switching. Through comparative simulations under varying channel occupancy rates, we delineated the conditions under which each model—NPCA and legacy—has a performance advantage. To address the limitations observed in both models, we proposed a hybrid model that assesses real-time channel occupancy and dynamically selects the optimal channel access method. This approach significantly enhances the overall throughput across a range of network conditions. Our study contributes valuable insights to the ongoing development of multi-channel access protocols within the IEEE 802.11 framework for the UHR group. The dynamic model introduced here presents a promising solution for improving network performance in environments with diverse and fluctuating traffic patterns.




\bibliographystyle{IEEEtran}
\bibliography{reference}
\end{document}